\documentclass[twocolumn,showpacs,preprintnumbers,amsmath,amssymb,prl,superscriptaddress]{revtex4-1}

\usepackage{graphicx}
\usepackage{color}
\usepackage[colorlinks=true,plainpages=false,linkcolor=blue,urlcolor=blue,citecolor=blue,pdfpagemode=UseNone,pdfstartview=FitBH]{hyperref}

\def\SIO{$\rm Sr_2IrO_4$}
\def\jeff{$j_{\rm{eff}}$}
\def\SLIO{Sr$_{2-x}$La$_x$IrO$_4$}

\begin{document}

\title{Persistent paramagnons deep in the metallic phase of Sr$_{2-x}$La$_x$IrO$_4$}

\author{H.~Gretarsson}

\author{N. H. Sung}
\author{J.~Porras}
\author{J. Bertinshaw}
\author{C. Dietl}
\author{Jan A. N. Bruin}
\author{A. F. Bangura}
\affiliation{Max-Planck-Institut f\"{u}r Festk\"{o}rperforschung, Heisenbergstr. 1, D-70569 Stuttgart, Germany}
\author{Y. K. Kim}
\affiliation{Advanced Light Source, Lawrence Berkeley National Laboratory, Berkeley, California 94720, USA}
\affiliation{Center for Correlated Electron Systems, Institute for Basic Science, Seoul 151-742, South Korea}
\affiliation{Department of Physics and Astronomy, Seoul National University, Seoul 151-747, South Korea}
\author{R. Dinnebier}
\affiliation{Max-Planck-Institut f\"{u}r Festk\"{o}rperforschung, Heisenbergstr. 1, D-70569 Stuttgart, Germany}
\author{Jungho~Kim}
\affiliation{Advanced Photon Source, Argonne National Laboratory,
Argonne, Illinois 60439, USA}
\author{A. Al-Zein }
\author{M. Moretti Sala}
\author{M. Krisch }
\affiliation{European Synchrotron Radiation Facility, BP 220, F-38043 Grenoble Cedex, France}
\author{M. Le Tacon}
\affiliation{Max-Planck-Institut f\"{u}r Festk\"{o}rperforschung, Heisenbergstr. 1, D-70569 Stuttgart, Germany}
\affiliation{Karlsruher Institut f\"{u}r Technologie, Institut f\"{u}r Festk\"{o}rperphysik, Hermann-v.-Helmholtz-Platz 1, D-76344 Eggenstein-Leopoldshafen, Germany}
\author{B. Keimer}
\affiliation{Max-Planck-Institut f\"{u}r Festk\"{o}rperforschung, Heisenbergstr. 1, D-70569 Stuttgart, Germany}
\author{B.~J.~Kim}
\affiliation{Max-Planck-Institut f\"{u}r Festk\"{o}rperforschung, Heisenbergstr. 1, D-70569 Stuttgart, Germany}

\date{\today}

\begin{abstract}

We have studied the magnetic excitations of electron-doped Sr$_{2-x}$La$_x$IrO$_4$ ($0 \leq x \leq 0.10$) using  resonant inelastic x-ray scattering (RIXS) at the Ir L$_3$-edge. The long range magnetic order is rapidly lost with increasing $x$, but two-dimensional short-range order (SRO) and dispersive magnon excitations with nearly undiminished spectral weight persist well into the metallic part of the phase diagram. The magnons in the SRO phase are heavily damped and exhibit anisotropic softening. Their dispersions are well described by a pseudospin-1/2 Heisenberg model with exchange interactions whose spatial range increases with doping. We also find a doping-independent high-energy magnetic continuum, which is not described by this model. The spin-orbit excitons arising from the pseudospin-3/2 manifold of the Ir ions broaden substantially in the SRO phase, but remain largely separated from the low-energy magnons. Pseudospin-1/2 models are therefore a good starting point for the theoretical description of the low-energy magnetic dynamics of doped iridates.

\end{abstract}


\maketitle

The proximity of two-dimensional antiferromagnetism and high-temperature superconductivity in copper oxides \cite{Keimer_Nature} and iron pnictides \cite{Paglione_Nature} suggest that both phases are intimately related. Neutron scattering \cite{Fujita_review_2012,Tranquada_review_2014} and resonant inelastic x-ray scattering (RIXS) \cite{Ament_RMP_RIXS_2011} experiments on both sets of materials \cite{LeTacon_NatPhys_2011,MPDean_NatMat_2013,Minola_PRL_2015,KIshii_NatComm_2013,WSLee_NatPhys_2014,KeJin_NatComm_2012} have indeed revealed  damped spin excitations in the superconducting regimes of the phase diagrams. In RIXS experiments, their dispersions and spectral weights are closely similar to those of magnons in the magnetically ordered parent compounds.

The notion of magnetically mediated high-temperature superconductivity has motivated an extensive search for new materials that realize two-dimensional quantum antiferromagnets akin to those in the undoped cuprates and iron pnictides. The antiferromagnetic Mott insulator Sr$_2$IrO$_4$ has emerged as a particularly promising candidate, because it is isostructural to La$_2$CuO$_4$, the progenitor of one of the most prominent families of superconducting cuprates, and it exhibits a closely analogous electronic structure. The Mott-insulating state in Sr$_2$IrO$_4$ is driven by the combination of intra-atomic spin-orbit coupling and electronic correlations (``spin-orbit Mott insulator''). Its magnetic ground state and low-energy magnetic excitations are well described by the pseudospin $j_{\rm{eff}}$=1/2 states arising from the spin-orbit coupled spin and orbital angular momenta of the iridium ions \cite{BJKim_PRL_2008,BJKim_Science_2009}. The pseudospins decorate a square lattice, and their interactions are well described by a Heisenberg model \cite{Jungho_PRL_2012} akin to the spin-1/2 Hamiltonian describing the magnetic dynamics of La$_2$CuO$_4$ \cite{Coldea_PRL_2001}.

These similarities are driving theoretical work predicting $d$-wave superconductivity in electron-doped Sr$_2$IrO$_4$ \cite{FaWang_PRL2011,Watanabe_PRL_2013,Yang_PRB2014,Kee_PRL_2014}. Evidence supporting this prediction has emerged from angle-resolved photoemission spectroscopy  (ARPES) measurements on electron-doped Sr$_2$IrO$_4$ surfaces, doped by charge transfer from a monolayer of adatoms. The Fermi surface of the surface electron system is split up into disjointed segments (``Fermi arcs'') \cite{Kim_Science2014} and exhibits a gap with $d$-wave symmetry at low temperature \cite{Kim_dWave,Feng_dWave} -- features that are hallmarks of $d$-wave superconductivity in the hole-doped cuprates \cite{Keimer_Nature}. These results have also been partially reproduced in \SLIO \cite{Baumberger_PRL_2015}, although impurity-driven disorder   \cite{Wilson_LaDoped_PRB_2015}  and oxygen vacancies \cite{Nakheon_arXiv_2015} have been an impediment to the progress in bulk, chemically-doped counterpart.

Despite these encouraging results, it is still unclear whether the magnetic correlations are as robust against carrier doping in Sr$_2$IrO$_4$ as they are in the cuprates. In particular, the on-site Coulomb repulsion is relatively weak in Sr$_2$IrO$_4$, resulting in a small charge gap ($\sim 0.4$ eV \cite{Moon_PRB_2009}) which is comparable to the magnon bandwidth \cite{Jungho_PRL_2012}. Recent Raman experiments have also revealed a strong pseudospin-lattice coupling in Sr$_2$IrO$_4$, indicating unquenched orbital dynamics that is quite unusual for a Mott insulator \cite{Gretarsson_TwoMagnon_2015}. These observations indicate a complex interplay between pseudospin, charge, and lattice degrees of freedom. All of these variables are expected to be sensitive to chemical doping, highlighting the need for a systematic study on how the magnetic correlations evolve in electron-doped Sr$_2$IrO$_4$.

\begin{figure}[htb]
\includegraphics[width=0.975\columnwidth]{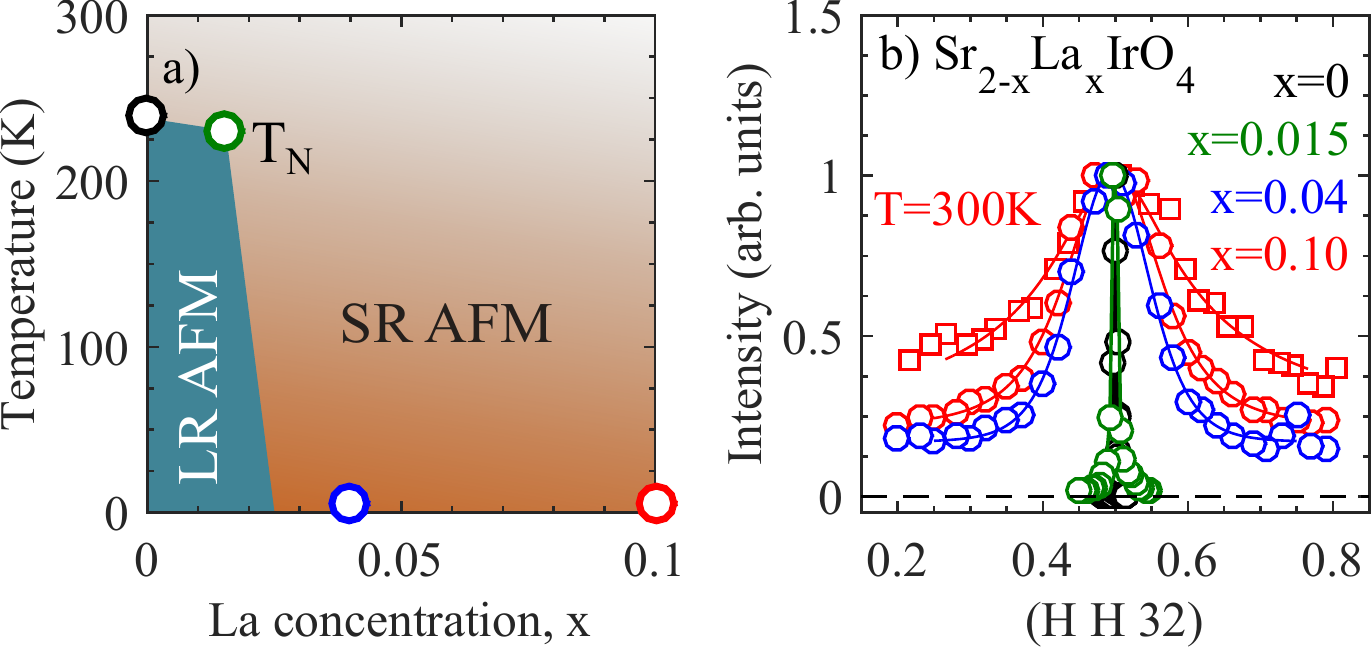}
\caption{\label{fig01}(Color online)  (a) Schematic phase diagram of Sr$_{2-x}$La$_x$IrO$_4$. 
Data points for the Neel temperature are taken from the magnetization measurements in Ref. \cite{sup}.   (b) $HH$-scans through the magnetic Bragg peak  Q$_{\rm M}=$ (0.5 0.5 $L_{even})$  at  $T = 20$ K (includes x=0.10 at $T=300$ K). The maximum intensity of each scan was normalized to unity. }
\end{figure}

Here we report a systematic investigation of the doping evolution of the magnetic structure and magnetic excitations in Sr$_{2-x}$La$_x$IrO$_4$ (x=0, 0.015, 0.04, and 0.10) by Ir L$_3$-edge RIXS. Our results show that antiferromagnetic long-range order (LRO) is rapidly lost upon electron doping ($0.015\!\!<\!\!x_c\!\!\leq$0.04), followed by persistent two-dimensional short-range order (SRO)  deep in the metallic phase of \SLIO. We report a detailed description of the low-energy magnon and high-energy spin-orbit exciton in the SRO phase, which provides an excellent basis for the theoretical description of the electronic properties of the doped iridates.

The RIXS experiments were carried out at the European Synchrotron Radiation Facility using the ID20 beamline. A spherical (2 m radius) diced Si(844) analyzer with a 60 mm mask and a Si(844) secondary monochromator were used to obtain an overall energy resolution of $\sim 25$~meV (full width at half maximum) and an in-plane momentum resolution of $\sim 0.035$ reciprocal lattice units. (The momentum transfer Q$=(H,K,L)$ is quoted in terms of reciprocal vector  $\pi/(a,b,c)$ where $a\approx b \approx 3.88$ ${\rm \AA}$ (undistorted unit cell) and $c \approx $ 25.8 ${\rm \AA}$ are the lattice parameters. In-plane crystal momenta q are quoted in square-lattice notation with unit lattice constant.) ARPES measurements were performed at Beamline 4.0.3 at the Advanced Light Source. The energy of the incident light was $h\nu$=90 eV and overall energy resolution of $\sim 25$~meV  was achieved.  Single crystals of La-doped \SIO were grown by the flux method, previously described in detail \cite{Nakheon_arXiv_2015}. The La concentrations were checked via electron probe micro-analysis. 

\begin{figure*}[htb]
\includegraphics[clip=true,  width=1.9\columnwidth]{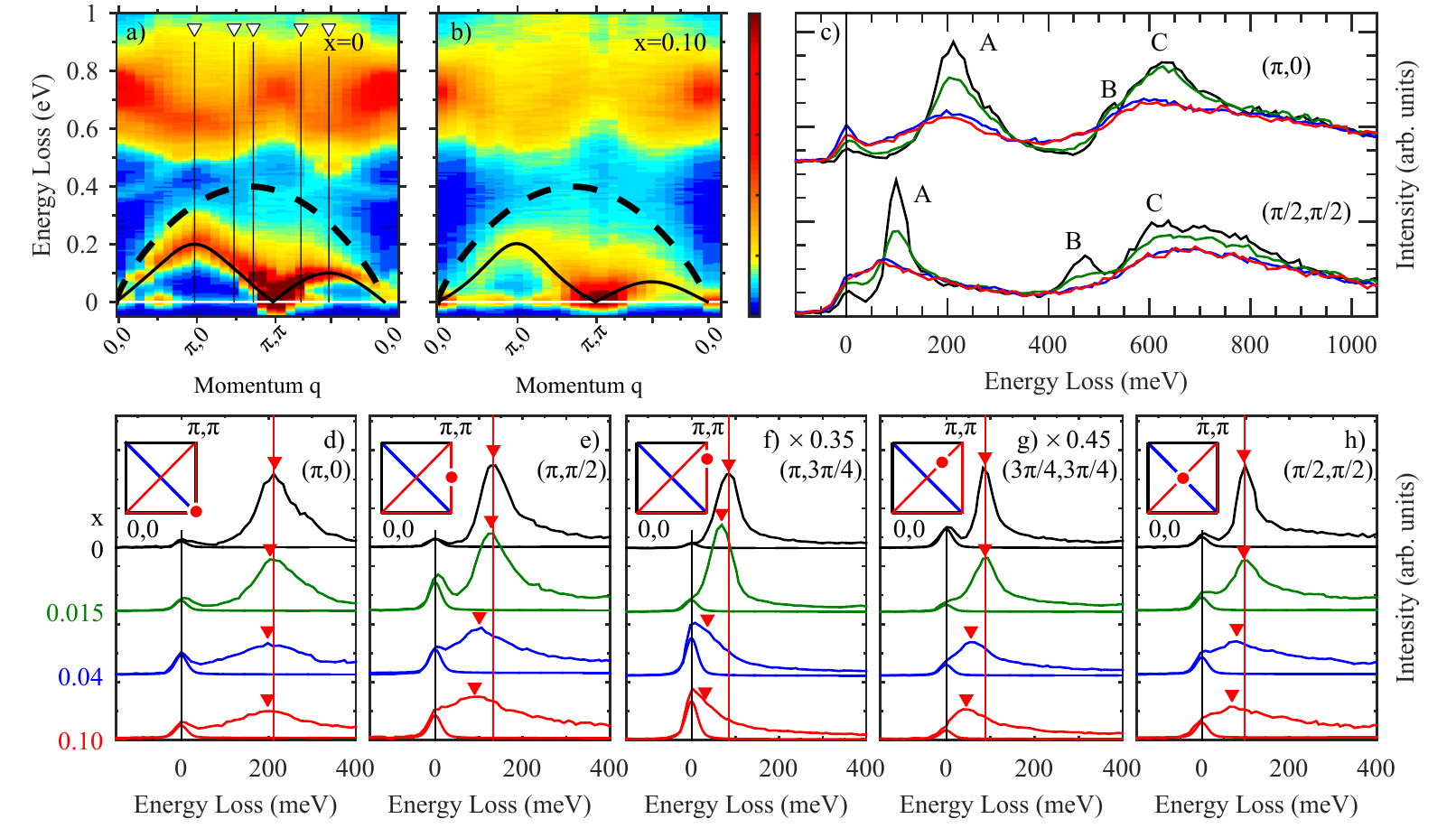}
\caption{\label{fig02}(Color online) RIXS intensity maps of \SLIO\ taken along high symmetry directions for (a) x=0 and (b) x=0.10. The intensity is in arbitrary units and starts from zero. Data were collected at $T = 20$ K. The spectral weight around 1 eV was used to normalize the RIXS spectra of different samples. The solid black lines are taken from the fit in Fig. \ref{fig03} and the dashed black line is a guide to eye, representing the upper boundary of the doping-independent continuum.  Constant-momentum cuts for all samples are displayed in (c) for q=($\pi,0$) and q=($\pi/2,\pi/2$). Letters refer to the excitations discussed in text. (d-h) Comparison of individual low-energy RIXS spectra for different doping levels and crystal momenta (see black vertical lines in (a)).  Red circles in insets show the q in Brillouin zone of the undistorted tetragonal unit cell (black square), the magnetic cell (blue line) is also shown. The contribution from elastic scattering has been superimposed on each spectrum. The red triangles represent the maximum intensity of the magnon modes (see text for details), and the vertical red line the magnon energy for the parent compound.}
\end{figure*}

The physical properties of Sr$_2$IrO$_4$ are very sensitive to the doping level and to the crystal growth conditions \cite{Cao_LaDoped_PRB_2011,Wilson_LaDoped_PRB_2015,Nakheon_arXiv_2015}. We have therefore carefully characterized all samples using susceptibility and resonant x-ray diffraction measurements.  Fig. \ref{fig01}(a) shows the phase diagram derived from our magnetization measurements (see  Supplemental Material (SM) in Ref. \cite{sup}). In crystals with $x=0$ and 0.015, the long range magnetic order (LRO)  \cite{BJKim_Science_2009} sets in around T$_N$ = 240 K and 230 K, respectively, while no signature of such order is found in  both $x=0.04$ and 0.10, indicating a doping-induced transition into a paramagnetic state. This observation is confirmed by monitoring the antiferromagnetic Bragg peaks, which are visible in the elastic scattering channel of the RIXS spectra for Q$_{\rm M}=$ (0.5 0.5 $L_{even})$ (Fig. \ref{fig01} (b)). Whereas in the $x=0$ and 0.015 crystals the magnetic Bragg peaks are sharp and narrow along the $HH$-directions, the elastic magnetic response of the $x=0.04$ and $x=0.10$ samples becomes extremely broad along the $HH$-direction and no correlation are observed along the $L$-direction (see SM in Ref. \cite{sup}), implying two-dimensional (2D) magnetic short-range order.  By fitting the profiles to 2D Lorentzians, we determined the in-plane correlation lengths $\xi = 10$ ${\rm\AA}$ and 8 ${\rm\AA}$ for $x=0.04$ and $x=0.10$, respectively \cite{note}. We also note that the 2D magnetic correlations persist to high temperature ($\xi \sim 5$ ${\rm\AA}$ for $x = 0.10$ at $T = 300$ K), consistent with the large in-plane exchange coupling determined by RIXS and Raman scattering \cite{Jungho_PRL_2012,Gretarsson_TwoMagnon_2015}.

The rapid suppression of magnetic LRO in electron-doped Sr$_{2-x}$La$_x$IrO$_4$ (see  Figure \ref{fig01}(a)) agrees with neutron diffraction results from Ref. \onlinecite{Wilson_LaDoped_PRB_2015} and parallels the behavior in hole-doped cuprates (such as La$_{2-x}$Sr$_{x}$CuO$_4$), where the LRO is quenched for $x \sim0.02$ \cite{Keimer_PRB_LCO_1992}. In the electron-doped cuprates, on the other hand, LRO survives well above 0.1 electrons per Cu \cite{Armitage_RMP_2010}. This supports the analogy between electron doping in the iridates and hole doping in the cuprates proposed on the basis of the opposite signs of the next-nearest-neighbor hopping amplitudes in the two sets of materials \cite{FaWang_PRL2011}.

Having studied the static magnetic properties, we now turn to the evolution of the magnetic excitations in \SLIO\ across the transition between LRO and SRO phases. Figure \ref{fig02} (a,b) displays color maps of RIXS spectra (x=0 and x=0.10)  taken at $T = 20$ K along high-symmetry directions of reciprocal space.  In agreement with prior work on the undoped compound \cite{Jungho_PRL_2012}, the data show a highly dispersive magnon excitation emanating from the antiferromagnetic ordering vector and extending up to about 0.2 eV. Our new data on the highest doped compound shows that magnon excitations (paramagnons) with closely similar dispersion and nearly undiminished spectral weight persist deep into the phase diagram of \SLIO. The magnetic dynamics in the doped iridates are thus strongly reminiscent of those in the hole-doped layered cuprates, where paramagnon excitations akin to the spin waves of antiferromagnetic La$_2$CuO$_4$ were found deeply in the metallic and superconducting regimes of the phase diagram \cite{LeTacon_NatPhys_2011,MPDean_NatMat_2013}.

\begin{figure}[htb]
\includegraphics[ width=0.95\columnwidth]{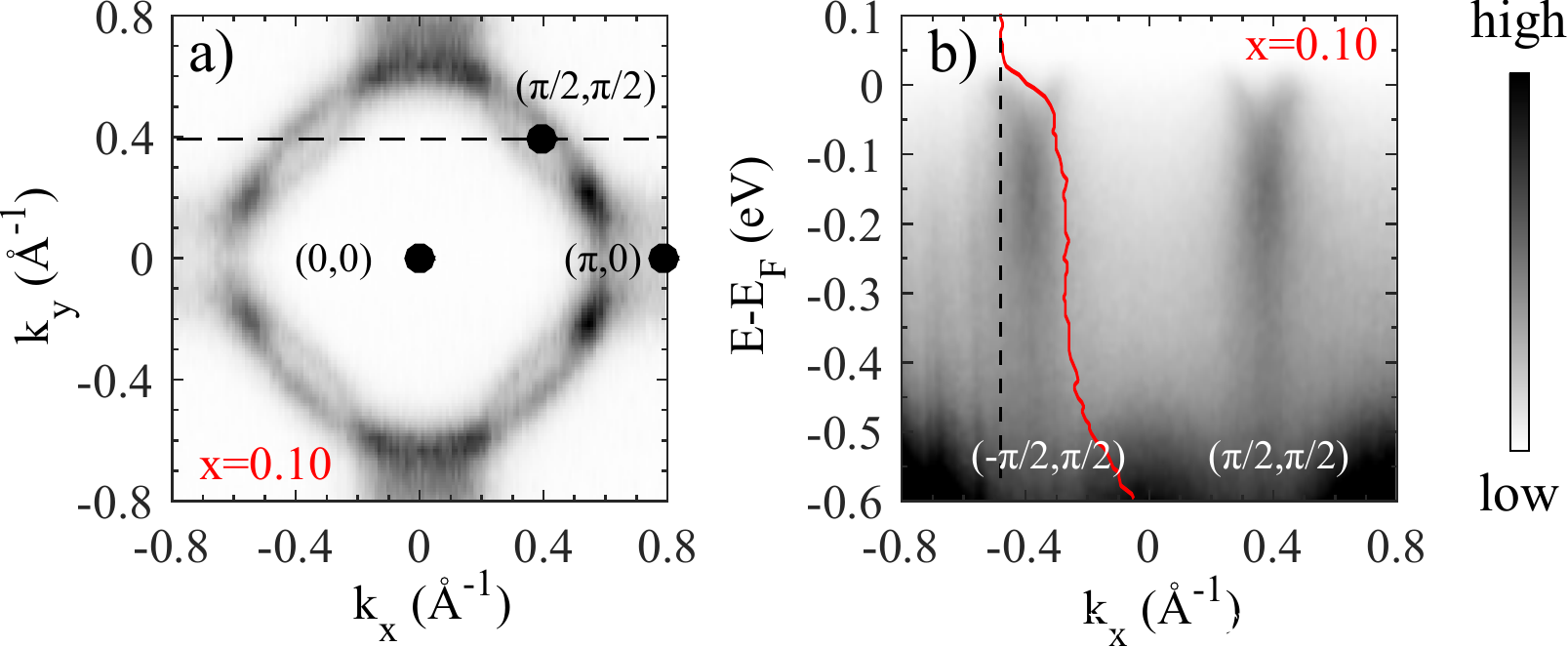}
\caption{\label{fig03}(Color online) (a) ARPES intensity map showing the Fermi surface of the x=0.10 sample. (b) Cut along the dashed horizontal line in (a). The red solid line shows the energy distribution curve at the momentum (vertical dashed line) where the dispersive charge excitations reach the Fermi level ($E_F$).}
\end{figure}

To emphasize the importance of our findings, we show in Fig. \ref{fig03} (a) an ARPES intensity map in the x=0.10 sample, taken at $T = 10$ K and at the Fermi level energy (E$_F$). Despite the detection of well-defined paramagnons in the x=0.10 sample a clear Fermi surface (FS) is apparent, indicating that the sample is deep in the metallic phase \cite{note2}.  The dispersive charge excitation, crossing E$_F$, can be seen in more detail in \ref{fig03} (b) where we plot the binding energy as a function of momentum (taken along the dashed horizontal line in (a)).  In a previous report \cite{Baumberger_PRL_2015}, it has been shown that this band is gapped near ($\pi,0$), making it a ``Fermi arc'' instead of a usual FS. Investigating how the magnetic correlations can coexist with a FS (i.e. system with Fermi-Dirac statistics), similar to the case in optimally hole-doped cuprates, is a question that should be addressed. One possible explanation for this discrepancy could be electronic phase separation on a macroscopic scale. This would give rise to a RIXS spectrum that looks like a superposition of metallic and insulating parts which is not supported by our data (e.g. Fig. \ref{fig02}(f) shows no hint of parent phase in the x=0.10 sample). Microscopic phase separation, as seen in STM measurements on metallic \SLIO\ \cite{Wilson_LaDoped_PRB_2015}, can however not be excluded.

\begin{figure}[htb]
\includegraphics[trim=5.25cm 15.85cm 5.25cm 6.5cm, clip=true,  width=0.95\columnwidth]{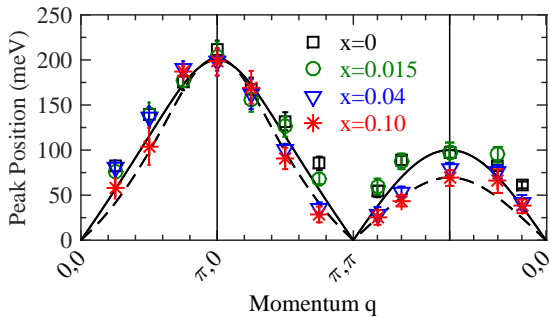}
\caption{\label{fig04}(Color online) Experimental magnon dispersion at $T = 20$ K for different doping levels. The solid (dashed) line represents the result of the best fit for $x=0$ and 0.015 ($x=0.04$ and 0.10), with parameters stated in the text.}
\end{figure}

To gain more insight into the magnetic excitations it is important to look systematically at how the RIXS spectra evolve with increasing doping.  Fig. \ref{fig02} (c) presents two salient features of the magnon (A) in the constant-momentum cuts plots (x=0, 0.015, 0.04 and x=0.10). First, the magnon energy at q=($\pi,0$) remains unchanged, although the magnon becomes severely damped. Second, at q=($\pi/2,\pi/2$), the magnon shows clear softening in addition to becoming broader.  The observed anisotropic magnetic softening along $(0,0)\rightarrow(\pi,\pi)$ direction is in stark contrast to the hardening of the magnon in electron-doped cuprates but resembles results seen in the metallic Bi-based cuprates \cite{Grioni_NatComm_2014,MPDean_PRB_2014}. This strengthens the analogy between electron-doped iridates and hole-doped cuprates already pointed out above.

To quantify the doping dependence of the magnon (or paramagnon) dispersion relations, we have tracked the peak positions in the RIXS spectra as a function of momentum (e.g. red triangles in Fig. \ref{fig02}(d-h)). The resulting dispersions are plotted in Fig. \ref{fig04}. The error bars were determined by selecting the energy interval in which the measured intensity is within 95$\%$ of the maximum intensity, following a recent RIXS study of a doped cuprate \cite{DJHuang_RIXS_2015}. Whereas elastic and inelastic contributions could be reliably separated in most of the spectra, the magnon peak positions for q=($0,0$) and q=($\pi,\pi$) could not be accurately determined due to the low intensity of the magnetic signal and the strong elastic line, respectively. These q-points were therefore not included in the dispersions in Fig. \ref{fig04}.

To obtain the doping dependent magnetic interactions, we have fitted the magnon dispersions obtained in this way by a Heisenberg model with exchange interactions between first, second, and third nearest neighbors in the IrO$_2$ planes, termed $J$, $J^{\prime}$, $J^{\prime\prime}$. For the $x=0$ and $x=0.015$ samples, we obtain $J=60$ meV, $J^\prime=-20$ meV, and $J^{\prime\prime}=15$ meV, in excellent agreement with prior work on undoped \SIO\ (solid lines in Fig. \ref{fig04} ) \cite{Jungho_PRL_2012}. To describe the anisotropic softening of the magnon dispersion in the $x=0.04$ and 0.10 samples noted above, the fit parameters were altered to $J=48$ meV, $J^\prime=-27$ meV, and $J^{\prime\prime}=20$ meV (dashed lines in Fig.  \ref{fig04}); here, the ratio between $J^{\prime}$ and $J^{\prime\prime}$ was fixed in order to reduce the free parameters. Doping $\rm Sr_2IrO_4$ with electrons thus reduces the nearest-neighbor exchange interaction by $\sim20\%$, while enhancing the longer-range interactions by $\sim30\%$. This finding is in general accord with expectations for magnetic interactions in itinerant-electron systems, and it will guide future theoretical work on electronic correlations in doped iridates.

The intensity maps in Fig. \ref{fig02}(a,b) and constant-momentum cuts in Fig.  \ref{fig02}(d-h) highlight yet another intriguing feature of this 2D quantum magnet.  By looking at the doping dependence at $(\pi/2,\pi/2)$, it becomes clear that the magnetic excitation has two components: the sharp mode discussed above which  reacts strongly to electron-doping, and a high-energy tail which retains its intensity upon doping (e.g. Fig. \ref{fig02}(c)). This momentum dependent high-energy tail is bounded at low energies by the magnon dispersion, and at high energies by the dashed black line in Fig. \ref{fig02}(a,b). The same behavior is also evident along the Brillouin zone border (e.g. Fig. \ref{fig02}(e)). One explanation for this marked departure in doping behavior between the single-magnon and the high-energy tail is that the latter arises from spinon-like excitations not directly associated with the N\'eel-ordered state. Indeed, in the cuprates, the spinon-picture \cite{Anderson_PRL_2001} has been evoked to explain the strong damping of the magnon at $(\pi,0)$ \cite{Perring_PRL_2010} and  the pronounced asymmetrical line shape of the two-magnon Raman scattering \cite{Suga_PRB_1990,Kampf_PRB_2003}. The latter was recently also observed in $\rm Sr_2IrO_4$ \cite{Gretarsson_TwoMagnon_2015}. Although this analogy and explanation are intriguing, further measurements, targeting the polarization dependence of the signal \cite{Ronnow_NatPhys2015}  are required to determine the origin of the high-energy magnetic tail in $\rm Sr_2IrO_4$.

For excitation energies above the optical gap ($> 0.4$ eV), we find a dispersive spin-orbit exciton mode arising from the \jeff=3/2 pseudospin manifold  \cite{Moon_PRB_2009,Jungho_PRL_2012}. In the parent compound, this feature comprises a sharp excitation near the optical gap (marked B in Fig. \ref{fig02}(c)) followed by incoherent higher-energy excitations (marked C). As pointed out in Ref. \onlinecite{Jungho_NatComm_2014}, the broadening of feature C may be due to overlap with the particle-hole continuum. Upon La-doping, the dispersion of the spin-orbit excitons (B+C) becomes less pronounced, and the sharp feature B disappears.  At ($\pi/2,\pi/2$), in particular, feature B is drastically suppressed upon doping, while C loses intensity. The strong damping of the spin-orbit exciton can be attributed to a lowering of the particle--hole continuum threshold as the system becomes metallic.

In summary, our systematic RIXS measurements of electron-doped Sr$_{2-x}$La$_x$IrO$_4$ have uncovered persistent short-range magnetic order deep in the metallic phase of \SLIO\, accompanied by dispersive magnetic excitations (paramagnons) of nearly undiminished spectral weight. Both the persistence of magnon-like excitations upon doping and their anisotropic softening indicate an intriguing analogy between electron-doped iridates and hole-doped cuprates. The low-energy magnon dispersions are well described by a $j_{\rm{eff}}$=1/2 Heisenberg model with exchange interactions whose spatial range increases with increasing doping. A doping independent high-energy magnetic continuum has also been found, suggesting that the ground state of \SIO\ contains additional short-range correlations not captured by the Heisenberg model. Finally, the $j_{\rm{eff}}$=3/2 spin-orbit exciton broadens and becomes less dispersive with increasing doping, but the overlap with the low-energy paramagnon branch remains small. Models based on $j_{\rm{eff}}$=1/2 pseudospins therefore appear to be a good starting point for the theoretical description of the electronic structure and for a realistic assessment of the prospects for unconventional superconductivity in doped iridates.

\textit{Note added}: Damped magnon and anisotropic softening have recently also been observed by Xuerong Liu and collaborators in La-doped \SIO ~\cite{Dean_arXiv_2016}.

\acknowledgements{We would like to thank G. Jackeli, G. Khaliullin, and R. Coldea for fruitful
discussions.  N. H. Sung was supported by the Alexander von
Humboldt Foundation. We acknowledge financial support by the DFG under grant TRR80.}


\begin{thebibliography}{33}

\bibitem{Keimer_Nature} B. Keimer, S. A. Kivelson, M. R. Norman, S. Uchida,  and J. Zaanen, \href{http://www.nature.com/nature/journal/v518/n7538/full/nature14165.html}{Nature {\bf 518}, 179 (2015)}.


\bibitem{Paglione_Nature} J. Paglione	and R. L. Greene, \href{http://www.nature.com/nphys/journal/v6/n9/abs/nphys1759.html}{Nature Physics {\bf 6}, 645 (2010)}.

\bibitem{Fujita_review_2012} M. Fujita, H. Hiraka,  M. Matsuda,  M. Matsuura,  J. M. Tranquada,  S. Wakimoto,  G. Xu,  K. Yamada, \href{ http://dx.doi.org/10.1143/JPSJ.81.011007}{Journal of the Physical Society of Japan {\bf 81}, 011007 (2012)}.

\bibitem{Tranquada_review_2014} J. M. Tranquada, G. Xu, I. A. Zaliznyak, \href{http://dx.doi.org/10.1016/j.jmmm.2013.09.029}{Journal of Magnetism and Magnetic Materials {\bf 350}, 148 (2014)}.

\bibitem{Ament_RMP_RIXS_2011} L. J. P. Ament, M. van Veenendaal, T. P. Devereaux, J. P. Hill, and J. van den Brink, \href{http://journals.aps.org/rmp/abstract/10.1103/RevModPhys.83.705} {Rev. Mod. Phys. {\bf 83}, 705 (2011)}.

\bibitem{LeTacon_NatPhys_2011} M. Le Tacon,	G. Ghiringhelli,	J. Chaloupka,	M. Moretti Sala,	V. Hinkov,	M. W. Haverkort,	M. Minola,	M. Bakr,	K. J. Zhou,	S. Blanco-Canosa,	C. Monney,	Y. T. Song,	G. L. Sun,	C. T. Lin, G. M. De Luca,	M. Salluzzo,	G. Khaliullin,	T. Schmitt,	L. Braicovich,	and  B. Keimer, \href{http://www.nature.com/nphys/journal/v7/n9/full/nphys2041.html} {Nature Physics {\bf 7}, 725-730 (2011)}.

\bibitem{MPDean_NatMat_2013} M. P. M. Dean,	G. Dellea,	R. S. Springell,	F. Yakhou-Harris,	K. Kummer,	N. B. Brookes,	X. Liu,	Y-J. Sun,	J. Strle,	T. Schmitt,	L. Braicovich,	G. Ghiringhelli,	I. Bo\v{z}ovi\'{c}, 	and J. P. Hill, \href{http://www.nature.com/nmat/journal/v12/n11/full/nmat3723.html} {Nature Materials {\bf 12}, 1019-1023 (2013)}.

\bibitem{Minola_PRL_2015} M. Minola, G. Dellea, H. Gretarsson, Y. Y. Peng, Y. Lu, J. Porras, T. Loew, F. Yakhou, N. B. Brookes, Y. B. Huang, J. Pelliciari, T. Schmitt, G. Ghiringhelli, B. Keimer, L. Braicovich, and M. Le Tacon, \href{http://journals.aps.org/prl/abstract/10.1103/PhysRevLett.114.217003} {Phys. Rev. Lett. {\bf 114}, 217003 (2015)}.

\bibitem{KIshii_NatComm_2013} K. Ishii,	M. Fujita,	T. Sasaki,	M. Minola,	G. Dellea,	C. Mazzoli,	K. Kummer,	G. Ghiringhelli,	L. Braicovich,	T. Tohyama,	 K. Tsutsumi,	K. Sato,	R. Kajimoto,	K. Ikeuchi,	K. Yamada,	M. Yoshida,	M. Kurooka,	and  J. Mizuki, \href{http://www.nature.com/ncomms/2014/140425/ncomms4714/full/ncomms4714.html} {Nature Communications {\bf 5}, 3714 (2014)}.

\bibitem{WSLee_NatPhys_2014} W. S. Lee,	J. J. Lee,	E. A. Nowadnick,	S. Gerber,	W. Tabis,	S. W. Huang,	V. N. Strocov,	E. M. Motoyama,	G. Yu,	B. Moritz,	H. Y. Huang,	R. P. Wang,	Y. B. Huang,	W. B. Wu,	C. T. Chen,	D. J. Huang,	M. Greven,	T. Schmitt,	Z. X. Shen,	and  T. P. Devereaux, \href{http://www.nature.com/nphys/journal/v10/n11/full/nphys3117.html} {Nature Physics {\bf 10}, 883-889 (2014)}.

\bibitem{KeJin_NatComm_2012} Ke-Jin Zhou,	Yao-Bo Huang,	C. Monney,	Xi Dai,	V. N. Strocov,	Nan-Lin Wang, Zhi-Guo Chen,	Chenglin Zhang,	Pengcheng Dai,	Luc Patthey,	J. van den Brink,	H. Ding,	and T. Schmitt, \href{http://www.nature.com/ncomms/journal/v4/n2/full/ncomms2428.html} {Nature Communications {\bf 4}, 1470 (2013)}.


\bibitem{BJKim_PRL_2008} B. J. Kim, Hosub Jin, S. J. Moon, J.-Y. Kim, B.-G. Park, C. S. Leem, Jaejun Yu, T. W. Noh, C. Kim, S.-J. Oh, J.-H. Park, V. Durairaj, G. Cao, and E. Rotenberg, \href{http://journals.aps.org/prl/abstract/10.1103/PhysRevLett.101.076402} {Phys. Rev. Lett. {\bf 101}, 076402 (2008)}.

\bibitem{BJKim_Science_2009} B. J. Kim, H. Ohsumi, T. Komesu, S. Sakai, T. Morita, H. Takagi, and T. Arima, \href{http://www.sciencemag.org/content/323/5919/1329} {Science {\bf 323}, 1329 (2009)}.


\bibitem{Jungho_PRL_2012} Jungho Kim, D. Casa,   M. H. Upton,  T. Gog,  Young-June Kim, J. F. Mitchell,  M. van Veenendaal,  M. Daghofer,  J. van den Brink, G. Khaliullin, and  B. J. Kim, \href{http://link.aps.org/doi/10.1103/PhysRevLett.108.177003} {Phys. Rev. Lett. {\bf 108}, 177003 (2012)}.


\bibitem{Coldea_PRL_2001} R. Coldea, S. M. Hayden, G. Aeppli, T. G. Perring, C. D. Frost, T. E. Mason, S.-W. Cheong, and Z. Fisk, \href{http://journals.aps.org/prl/abstract/10.1103/PhysRevLett.86.5377} {Phys. Rev. Lett. {\bf 86}, 5377 (2001)}.


\bibitem{FaWang_PRL2011} F. Wang and T. Senthil, \href{http://journals.aps.org/prl/abstract/10.1103/PhysRevLett.106.136402}{Phys. Rev. Lett. {\bf 106}, 136402 (2011)}.

\bibitem{Yang_PRB2014} Y. Yang, W.-S. Wang, J.-G. Liu, H. Chen, J.-H. Dai, and
Q.-H. Wang, \href{http://journals.aps.org/prb/abstract/10.1103/PhysRevB.89.094518}{Phys. Rev. B {\bf 89}, 094518 (2014)}.

\bibitem{Watanabe_PRL_2013} Hiroshi Watanabe, Tomonori Shirakawa, and Seiji Yunoki, \href{http://journals.aps.org/prl/abstract/10.1103/PhysRevLett.110.027002}{Phys. Rev. Lett. {\bf 110}, 027002 (2013)}.

\bibitem{Kee_PRL_2014} Zi Yang Meng, Yong Baek Kim, and Hae-Young Kee, \href{http://journals.aps.org/prl/abstract/10.1103/PhysRevLett.113.177003}{Phys. Rev. Lett. {\bf 113}, 177003 (2014)}.

\bibitem{Kim_Science2014} Y. K. Kim, O. Krupin, J. D. Denlinger, A. Bostwick, E. Rotenberg, Q. Zhao, J. F. Mitchell, J. W. Allen, and B. J. Kim, \href{http://www.sciencemag.org/content/345/6193/187.abstract}{Science {\bf 345}, 187 (2014)}.

\bibitem{Kim_dWave} Y. K. Kim, N. H. Sung, J. D. Denlinger, and B. J. Kim, \href{http://www.nature.com/nphys/journal/v12/n1/full/nphys3503.html}{Nature Physics {\bf 12}, 37 (2016)}.

\bibitem{Feng_dWave} Y. J. Yan, M. Q. Ren, H. C. Xu, B. P. Xie, R. Tao, H. Y.
Choi, N. Lee, Y. J. Choi, T. Zhang, and D. L. Feng, \href{http://journals.aps.org/prx/abstract/10.1103/PhysRevX.5.041018}{Phys. Rev. X {\bf 5}, 041018 (2015)}.

\bibitem{Wilson_LaDoped_PRB_2015} Xiang Chen, Tom Hogan, D. Walkup, Wenwen Zhou, M. Pokharel, Mengliang Yao, Wei Tian, Thomas Z.
Ward, Y. Zhao, D. Parshall, C. Opeil, J. W. Lynn, Vidya Madhavan, and Stephen D. Wilson, \href{http://journals.aps.org/prb/abstract/10.1103/PhysRevB.92.075125} {Phys. Rev. B {\bf 92}, 075125 (2015)}.


\bibitem{Nakheon_arXiv_2015} N. H. Sung, H. Gretarsson, D. Pr\"{o}pper, J. Porras, M. Le Tacon, A. V. Boris, B. Keimer, and B. J. Kim, \href{http://www.tandfonline.com/doi/abs/10.1080/14786435.2015.1134835} {Phil. Mag. {\bf 96}, 413 (2016)}.

\bibitem{Baumberger_PRL_2015} A. de la Torre, S. McKeown Walker, F. Y. Bruno, S. Ricco, Z. Wang, I. Gutierrez Lezama, G. Scheerer, G. Giriat, D. Jaccard, C. Berthod, T. K. Kim, M. Hoesch, E. C. Hunter, R. S. Perry, A. Tamai, and F. Baumberger, \href{http://journals.aps.org/prl/abstract/10.1103/PhysRevLett.115.176402}{Phys. Rev. Lett. {\bf 115}, 176402 (2015)}.

\bibitem{Moon_PRB_2009} S. J. Moon, Hosub Jin, W. S. Choi, J. S. Lee, S. S. A. Seo, J. Yu, G. Cao, T. W. Noh, and Y. S. Lee, \href{http://journals.aps.org/prb/abstract/10.1103/PhysRevB.80.195110} {Phys. Rev. B {\bf 80}, 195110 (2009)}.

\bibitem{Gretarsson_TwoMagnon_2015} H. Gretarsson, N. H. Sung, M. H\"{o}ppner, B. J. Kim, B. Keimer, and M. Le Tacon, \href{http://arxiv.org/abs/1509.03396} {arXiv:1509.03396 (2015)}.


\bibitem{Cao_LaDoped_PRB_2011} M. Ge, T. F. Qi, O. B. Korneta, D. E. De Long, P. Schlottmann, W. P. Crummett, and G. Cao, \href{http://journals.aps.org/prb/abstract/10.1103/PhysRevB.84.100402} {Phys. Rev. B {\bf 84}, 100402(R) (2011)}.


\bibitem{note} We cannot rule out that the broadening of the Bragg reflections is partially due to unresolved incommensurability of the magnetic correlations, in analogy to the hole-doped cuprates. The values of the correlation length we quote should therefore be regarded as lower bounds.

\bibitem{note2} A weakly metallic behavior is seen in  transport measurement on the x=0.04 sample \cite{sup}, indicating that the insulator to metal phase boundary occurs close to that doping level.


\bibitem{Keimer_PRB_LCO_1992} B. Keimer, N. Belk, R. J. Birgeneau, A. Cassanho, C. Y. Chen, M. Greven, M. A. Kastner, A. Aharony, Y. Endoh, R. W. Erwin, and G. Shirane, \href{http://journals.aps.org/prb/abstract/10.1103/PhysRevB.46.14034} {Phys. Rev. B {\bf 46}, 14034 (1992)}.



\bibitem{Armitage_RMP_2010}  N. P. Armitage,  P.  Fournier, and  R. L. Greene, \href{http://link.aps.org/doi/10.1103/RevModPhys.82.2421} {Rev. Mod. Phys. {\bf 82}, 2421 (2010)}.


\bibitem{Grioni_NatComm_2014} M. Guarise, B. Dalla Piazza, H. Berger, E. Giannini, T. Schmitt, H. M. R\o nnow,
G.A. Sawatzky, J. van den Brink, D. Altenfeld, I. Eremin, and M. Grioni, \href{http://www.nature.com/ncomms/2014/141218/ncomms6760/full/ncomms6760.html} {Nat. Comm. {\bf 5}, 5760 (2014)}.


\bibitem{MPDean_PRB_2014} M. P. M. Dean, A. J. A. James, A. C. Walters, V. Bisogni, I. Jarrige, M. H\"{u}cker, E. Giannini, M. Fujita, J. Pelliciari, Y. B. Huang, R. M. Konik, T. Schmitt, and J. P. Hill, \href{http://journals.aps.org/prb/abstract/10.1103/PhysRevB.90.220506} {Phys. Rev. B {\bf 90}, 220506 (2014)}.

\bibitem{DJHuang_RIXS_2015} H. Y. Huang, C. J. Jia, Z. Y. Chen, K. Wohlfeld, B. Moritz, T. P. Devereaux, W. B. Wu, J. Okamoto, W. S. Lee, M. Hashimoto, Y. He, Z. X. Shen, Y. Yoshida, H. Eisaki, C. Y. Mou, C. T. Chen, and D. J. Huang, \href{http://dx.doi.org/10.1038/srep19657}{Scientific Reports {\bf 6}, 19657 (2016)}.




\bibitem{Anderson_PRL_2001} Chang-Ming Ho, V. N. Muthukumar, Masao Ogata, and P. W. Anderson, \href{http://journals.aps.org/prl/abstract/10.1103/PhysRevLett.86.1626} {Phys. Rev. Lett. {\bf 86}, 1626 (2001)}.

\bibitem{Perring_PRL_2010} N. S. Headings, S. M. Hayden, R. Coldea, and T. G. Perring, \href{http://journals.aps.org/prl/abstract/10.1103/PhysRevLett.105.247001} {Phys. Rev. Lett. {\bf 105}, 247001 (2010)}.

\bibitem{Suga_PRB_1990} S. Sugai, M. Sato, T. Kobayashi, J. Akimitsu, T. Ito, H. Takagi, S. Uchida, S. Hosoya, T. Kajitani, and T. Fukuda, \href{http://journals.aps.org/prb/abstract/10.1103/PhysRevB.42.1045} {Phys. Rev. B {\bf 42}, 1045 (1990)}.


\bibitem{Kampf_PRB_2003} A. A. Katanin and A. P. Kampf, \href{http://journals.aps.org/prb/abstract/10.1103/PhysRevB.67.100404} {Phys. Rev. B {\bf 67}, 100404 (2003)}.

\bibitem{Ronnow_NatPhys2015}  B. Dalla Piazza,	M. Mourigal,	N. B. Christensen,	G. J. Nilsen,	P. Tregenna-Piggott,	T. G. Perring,	M. Enderle,	D. F. McMorrow,	D. A. Ivanov,	and  H. M. R\o nnow, \href{http://www.nature.com/nphys/journal/v11/n1/full/nphys3172.html} {Nature Physics {\bf 11}, 62 (2015)}.


\bibitem{Jungho_NatComm_2014} Jungho Kim,	M. Daghofer,	A. H. Said,	T. Gog,	J. van den Brink,	G. Khaliullin,	and  B. J. Kim, \href{http://www.nature.com/ncomms/2014/140717/ncomms5453/full/ncomms5453.html} {Nature Communications {\bf 5}, 4453 (2014)}.





\bibitem{sup} See Supplemental Material at http://link.aps.org/supplemental/ for additional information.

\bibitem{Dean_arXiv_2016} Xuerong Liu, M. P. M. Dean, Z. Y. Meng, M. H. Upton, T. Qi, T. Gog, H. Ding, G. Cao, and J. P. Hill,  \href{http://journals.aps.org/prb/abstract/10.1103/PhysRevB.93.241102} {Phys. Rev. B {\bf 93},  241102(R) (2016)}.




\end{thebibliography}
\end{document}